\documentclass[runningheads]{llncs}
\usepackage[T1]{fontenc}
\usepackage{graphicx}
\usepackage{amsfonts}
\usepackage{amsmath}
\usepackage{multirow}
\begin{document}
\title{Analyzing Brain Tumor Connectomics using Graphs and Persistent Homology}

\author{Debanjali Bhattacharya, Ninad Aithal, Manish Jayswal and Neelam Sinha}
\institute{Center for Brain Research, Indian Institute of Science, Bangalore}
\maketitle              
\begin{abstract}
Recent advances in molecular and genetic research have identified a diverse range of brain tumor sub-types, shedding light on differences in their molecular mechanisms, heterogeneity, and origins. The present study performs whole-brain connectome analysis using diffusion-weighted images. To achieve this, both graph theory and persistent homology—a prominent approach in topological data analysis are employed in order to quantify changes in the structural connectivity of the whole-brain connectome in subjects with brain tumors.
Probabilistic tractography is used to map the number of streamlines connecting 84 distinct brain regions, as delineated by the Desikan-Killiany atlas from FreeSurfer. These streamline mappings form the connectome matrix, on which persistent homology based analysis and graph theoretical analysis are executed to evaluate the discriminatory power between tumor sub-types that include meningioma and glioma. A detailed statistical analysis is conducted on persistent homology-derived topological features and graphical features to identify the brain regions where differences between study groups are statistically significant ($p<0.05$).
For classification purpose, graph-based local features are utilized, achieving a highest accuracy of 88\%. In classifying tumor sub-types, an accuracy of 80\% is attained. The findings obtained from this study underscore the potential of persistent homology and graph theoretical analysis of the whole-brain connectome in detecting alterations in structural connectivity patterns specific to different types of brain tumors.

\end{abstract}
\section{Introduction}
\label{intro}

Diffusion-weighted magnetic resonance imaging (DWI) tractography is a revolutionary imaging technique that allows for non-invasive reconstruction of the brain's white matter (WM) fibre tracks at macro scale \cite{ref1}. By tracing the diffusion of water molecules along axonal pathways, tractography can map the connections between different brain regions, providing insights into the structural connectivity of brain. This method has been significantly contributed across various neurological contexts, including aging, development, and disease characterization \cite{ref3,ref4}. Although tractography primarily employed a qualitative approach in visualizing WM connections, recent trends have seen the rise of quantitative methodologies, offering powerful tools for analyzing brain connectivity and tissue micro-structure through tractography. In this paper, we elucidate how tractography serves as a cornerstone for quantitative analysis in assessing changes of the brain structural connectivity for brain tumor classification problem.
Recent advances in genome-based molecular investigations and clinical studies have revealed different sub-types of brain tumor.
While numerous studies have explored various imaging features and their correlations with pathological and molecular characteristics, especially for tumor segmentation and classifying different grades of glioma, limited research has investigated the changes in structural connectivity of whole-brain due to tumor. Tumor often disrupt the normal architecture of brain by infiltrating healthy tissue, leading to changes in their connectivity patterns. Since, DWI tractography allows for visualization of WM tracts in the brain, in the context of studying brain tumor, tractography offers a unique opportunity to examine the changes in structural connectivity due to tumor with the surrounding brain tissues. The present study conducts a comprehensive topological data analysis and graph theoretical analysis on DWI whole-brain connectome matrix to differentiate between 3 study groups: (i) glioma, (ii) meningioma and (iii) healthy control.
The main purpose is to distinguish these two types of brain tumors having distinct origins, characteristics, and treatment approaches. Meningiomas originate from meninges, the protective membranes surrounding brain and spinal cord, while gliomas arise from glial cells, which provide support and protection to neurons in brain. Differentiating between meningioma and glioma is important for diagnosis and treatment planning, as treatment strategies and prognoses differ between these two tumor sub-types. Meningiomas are typically benign, slow-growing which typically exhibit more localized growth and minimal invasion into surrounding tissue, thus, often amenable to surgical resection with favorable outcomes. In contrast, gliomas originates from glial cells are characterized by infiltrative growth and can vary widely in aggressiveness and prognosis, necessitating a more nuanced approach to treatment, including surgery, radiation therapy, and chemotherapy. 

\begin{figure}
    \centering
    \includegraphics[width=\textwidth]{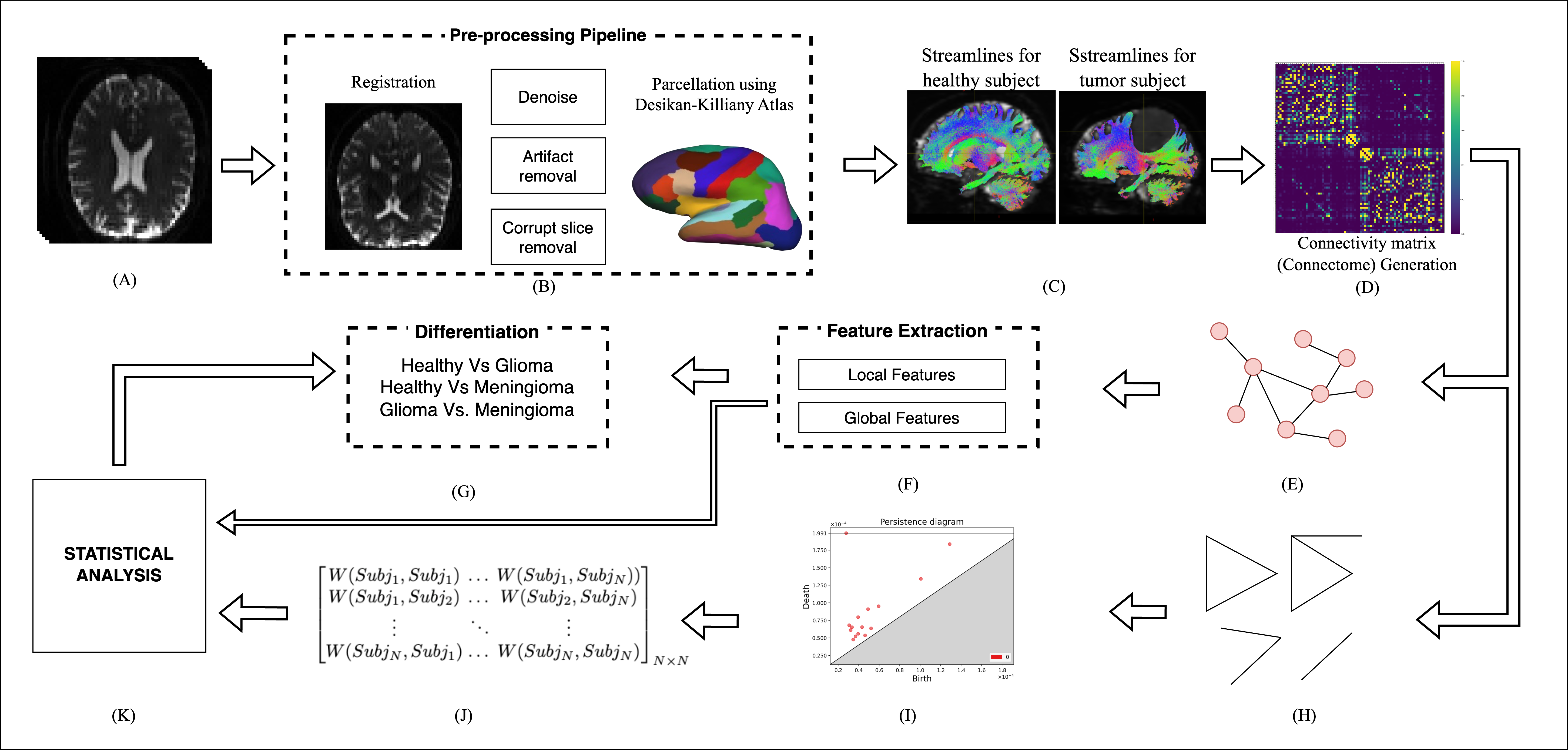}
    \caption{Block schematic of the proposed study}
    \label{fig:blockDiagram}
\end{figure}

In the current study, we focus on applying a biphasic approach to differentiate meningiomas and gliomas using DWI brain connectome.
Initially, we utilize persistent homology to capture the differences in topological characteristics of the brain connectome between the study groups. Persistent homology is crucial for identifying distinct topological features that might be overlooked by traditional graph measures, providing a deeper insight into the topological structure and its alterations due to different tumor types. This method enhances our ability to distinguish between the study groups by uncovering unique topological traits.
Additionally, our study employs DWI connectome-based graph measures to classify the study groups. The importance of computing graphical features lies in their ability to capture the subtle local alterations in brain structural connectivity induced by various types of tumors. This dual approach offers a comprehensive analysis in examining the structural differences in brain connectomes among healthy individuals, and patients with meningioma and glioma, making this study a significant step forward in the research of connectome-based analysis of brain tumors. The block schematic of the proposed methodology is shown in Figure~\ref{fig:blockDiagram}.



\section{Materials and Methods}

\subsection{Dataset and MRI preprocessing}
Publicly available Brain Tumor Connectomics (BTC) Dataset \cite{ref5} is utilized that include pre-operative data of 25 patients who were diagnosed with (i) glioma ($n=11$, mean age=47.5y, SD=11.3; M:F= 7:4), developing from glial cells, and (ii) meningioma ($n=14$, mean age=60.4y, SD=12.3; M:F=3:11), developing in the meninges. Patients were recruited from Ghent University Hospital (Belgium). A total 11 healthy control subjects (mean age=58.6y, SD=10.3; M:F= 7:4) are also incorporated for comparison purpose.
From all participants, two types of MRI scans are utilized using a Siemens 3T Magnetom Trio MRI scanner with a 32-channel head coil. These includes (i) T1-W structural MRI (160 slices, TR=1750 ms, TE=4.18 ms, voxel size $1\times1\times 1 mm$), (ii) a multishell high-angular resolution diffusion-weighted MRI (DWI) scan was acquired (60 slices; TR=8700 ms; TE=110 ms; 101 diffusion directions; b-values=$0, 700, 1200, 2800 s/mm^{2}$; voxel size =$2.5\times2.5\times 2.5 mm$). In addition, two DWI $b=0 s/mm^{2}$ images are used with reversed phase-encoding blips for the purpose of correcting susceptibility-induced distortions. Further details regarding characteristics of the participants and MRI data acquisition can be found in paper by Hannelore Aerts et.al \cite{ref5}.

\subsection{MRI preprocessing}

Structural MRI preprocessing utilized the automated FreeSurfer pipeline (version 7.1.0), which parcellated the cerebral cortex into gyral and sulcal structures based on 84 regions of interest (ROIs) from the Desikan-Killiany atlas (DK atlas) \cite{ref6}. The labels of DK ROIs are listed in the github repository\footnote{Link to code: https://github.com/blackpearl006/TGI3-2024/}. The preprocessing steps are conducted using MRtrix toolbox \cite{ref8} for DWI denoising, Gibbs’ ringing artifacts removal. Additionally, FSL (version 6.0.7.7) libraries are also used for preprocessing. It includes brain extraction with the brain extraction tool, motion correction, correction for susceptibility-induced distortion using topup and correction for eddy-current distortion.

\subsection{Whole-brain probabilistic tractography}

Anatomically-constrained probabilistic tractography \cite{ref7} is conducted using the MRtrix3 software package \cite{ref8}. Initially, five-tissue-type segmentation is performed, delineating cortical gray matter, subcortical gray matter, WM, cerebrospinal fluid, and other pathological tissues. Subsequently, constrained spherical deconvolution algorithm is used to estimate voxel-wise fibre orientation densities for each tissue type, providing insight into diffusion characteristics along multiple directions within a voxel. Since, MRtrix allows for the estimation of multiple crossing fibres within a single voxel, it can effectively disentangle the diffusion signal into multiple directions.
Finally, whole-brain tractography is executed, initiating from the voxels at the gray matter-WM interface as seed points: the initial positions from which streamlines representing WM pathways are propagated throughout the brain. For each subject, a total of 10,000,000 streamlines are generated. The default values for the maximum length of streamlines and the angular threshold are set to 250mm and 0.06 radians, respectively, that determine when streamline propagation terminates. These parameters collectively influence several aspects of tractography, encompassing seeding, propagation, and termination criteria, thereby aiding in the delineation of WM pathways in the brain. Following the creation of the streamline map, a \textit{connectome matrix} is generated, illustrating the number of streamlines connecting distinct brain regions (in this case, 84 ROIs of the DK atlas). However, computing DWI connectomes using MRtrix software is computationally very expensive, necessitating the use of high-performance computing (HPC) resources. Specifically, we employed an Intel(R) Xeon(R) Gold 6240 CPU @ 2.60GHz with dual CPUs and 192 GB of memory for this purpose.

\begin{figure}
  \centering
  \includegraphics[width=0.7\textwidth]{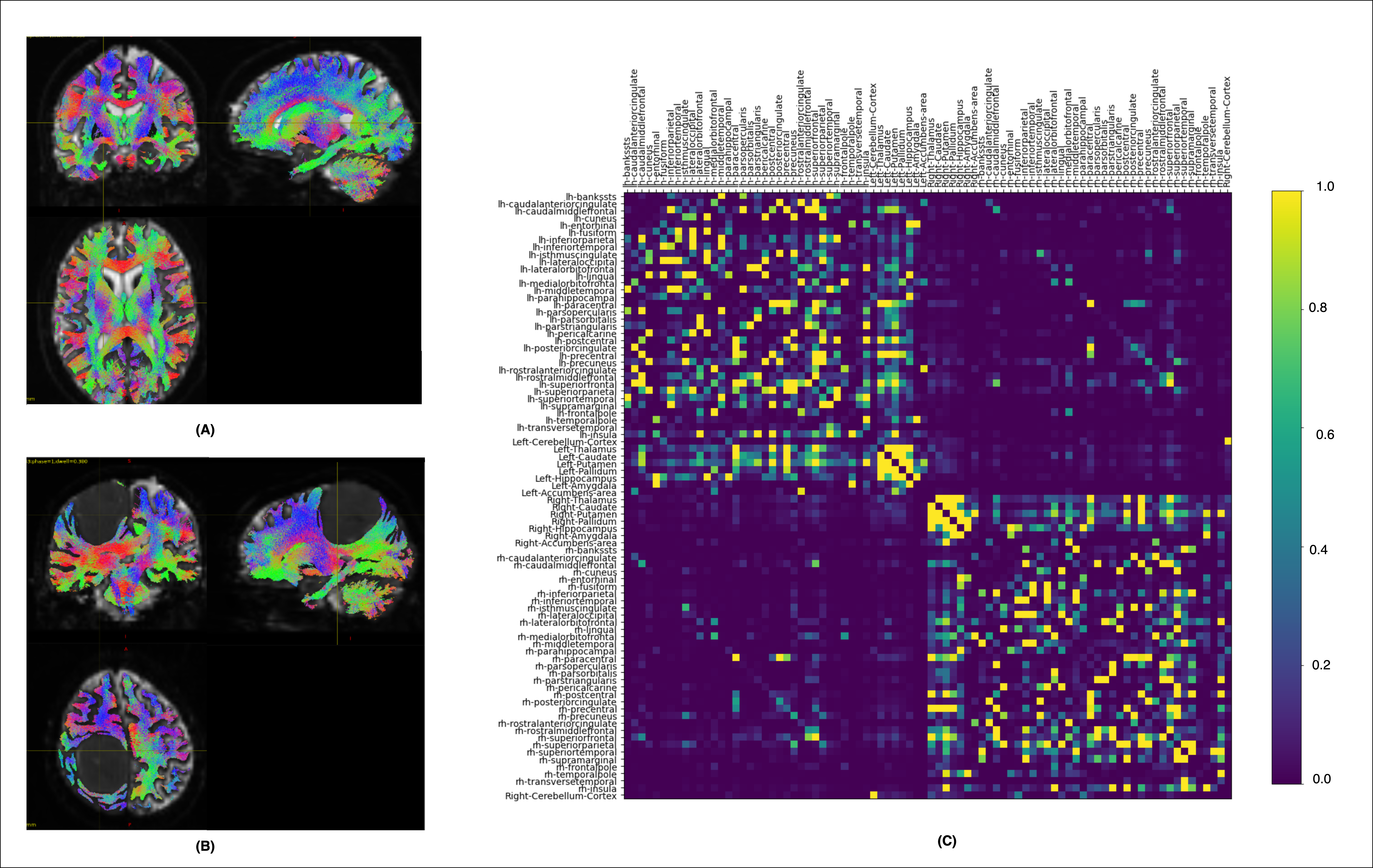}
 \caption{Visualization of generated streamlines is showed in three MRI planes for one healthy subject: Fig.(A) and one subject diagnosed with meningioma: Fig.(B). As seen from Fig.(B), no streamlines are observed within the region affected by the meningioma. Fig.(C) shows the resultant connectome as obtained from streamlines of one healthy subject. Enhanced structural connectivity within each hemisphere is evident, represented by the prominent diagonal boxes. Increased structural connectivity between homologous regions of the left and right hemisphere is depicted by brighter spots in the off-diagonal boxes.
} \label{fig:streamlines}
\end{figure}

\subsection{Persistent homology on whole-brain connectome}

Persistent homology is a powerful topological data analysis approach that provides a robust framework for analyzing topological features of data, particularly in the context of shape and structure. At its core, persistent homology aims to capture the evolution of topological features across different spatial scales by constructing a sequence of topological spaces based on the input data. It reveals how specific topological characteristics evolve as we observe these spaces at different levels of detail. 
Persistent homology finds its application various domains. In the context of medical image analysis, persistent homology is used for analysis of endoscopy \cite{2016-Dunaeva-Endoscopy-Analysis},  analysis of brain networks for differentiating various types of brain disorders \cite{2012-Lee-Persistent-Homology-Brain-Networks,2021-Stolz-schizophrenia-experiments}, analysis of visual brain networks \cite{boldfmrits}, and detecting transition between states in EEG \cite{2016-Merelli-EEG}, identifying epileptic seizures \cite{2021-Caputi-Epilepsy} and distinguishing between male and female brain networks \cite{2022-Das-TDA-Brain-Networks}. 
In this study, we exploit the information encoded in \textit{persistence diagram} to analyze DWI brain connectome of healthy individuals and patients having meningioma and glioma. Here, we provide an intuitive explanation of how persistence diagrams are constructed based on graph filtrations. Further details on persistence diagram can be found in literature \cite{2010-Edelsbrunner-book}.

To capture the $0$-dimensional homological features, the birth and death of connected components are examined in the sublevel set filtration. In computational topology, connected components are called $0$-dimensional homological features. Hence, the persistence diagram that encodes the birth and death of connected components is called the $0$-th ordinary persistence diagram, denoted by $H_{0}$. Mathematically, the points in $H_{0}$ are computed from the birth and death of homology classes in a sequence of homology groups corresponding to the sublevel set filtration \cite{2010-Edelsbrunner-book}. 
To capture the $1$-dimensional features (loops), we compute the dimension-1 extended persistence diagram. This diagram incorporates information from both the sublevel set and superlevel set filtrations \cite{2010-Edelsbrunner-book}. So, in case of $1$-st extended persistence diagram, in order to encode loops, one considers the birth and death of homology classes in a filtration obtained by combining the sublevel set and superlevel set filtrations, called extended filtration. Therefore, the persistence diagram encoding the loops of graph $G$ is called the $1$-st extended persistence diagram and is denoted by $H_{1}$. To construct $H_{1}$, one considers the $1$-dimensional homology classes that persist throughout the sequence of absolute homology groups corresponding to the sublevel set filtration. Such homology classes are called essential homology classes. Then, the deaths of these essential homology classes in the relative homology groups in the superlevel set filtration are considered. Note that, an essential homology class which is born at the sublevel set filtration and dies at the superlevel set filtration is encoded as a point in $H_{1}$. \par
The quantification of persistence diagrams utilizes the \textit{Wasserstein distance}, a metric that determines the dissimilarity (contributed by higher distance values) between two persistence diagrams. Specifically, it measures the extent of transformation needed for one persistence diagram to closely match another, providing a metric for comparing the topological structures of DWI connectome metrices. This distance metric captures the subtle variations and structural differences in brain connectivity patterns associated with different types of brain tumors. Thus, it serves as a key tool for understanding and quantifying the complex topological characteristics and their alterations in brain networks. 
In this study, Wasserstein distance is computed across all considered subjects for each of the Betti descriptors in dimension-0 ($H_0$) and dimension-1 ($H_1$). 
To compute the distance elements of two persistence diagrams $X$ and $Y$ one-to-one (bijection $\eta$) are matched. Mathematically, it is defined as,

\begin{equation}
W_{q,p}(X,Y) = \left[ \inf_{\eta:X\rightarrow Y} \sum_{x \in X} || x - \eta(x) ||_{\infty}^{q} \right]^{\frac{1}{q}}
\label{eq:wass}
\end{equation}

\subsection{Graph-based feature extraction}

The DWI whole-brain connectome matrices are also used to compute features based on graph theory, with edges indicating the strengths of fiber connectivity between all pairs of 84 ROIs (nodes). The Brain Connectivity Toolbox from MATLAB is utilized for this purpose \cite{graphpaperref22}. Here, we delve into graph-based characteristics that unveil both the functional integration and segregation aspects of structural connectivity, with the goal of understanding the changes in the individual brain regions in cases of glioma. Global measures of a graph provide insights into the functional segregation (the degree to which network elements form specific modules or clusters) and functional integration (the ability to combine segregated information from different modules of the network) of information flow within the brain network \cite{graphpaperref1,graphpaperref22}. Characteristic path length, a widely used measure, quantifies the properties of functional integration. Conversely, modularity and clustering coefficients are commonly used measures that quantify global information segregation in brain networks \cite{graphpaperref22}. Local features, on the other hand, aim to reveal the characteristics of each ROI in a brain network. 
In the present study, 4 global features - \textit{transitivity, modularity, characteristic path length, and density} - and 9 local features - \textit{clustering coefficient (CC), nodal degree (Deg), betweenness centrality (BC), local efficiency (LE), eigenvector centrality (EVC), participation coefficient (PC), diversity coefficient (DC), gateway coefficient (GC), and strength (Str)}, are extracted. Detailed descriptions of all the considered 13 features can be found in paper \cite{graphpaperref22}.

\section{Results}

The present study performs DWI brain connectome analysis for classifying brain tumor with different origins. Probabilistic tractography is used to generate the streamlines which are the WM tracks connecting 84 different regions of gray matter as defined by DK atlas. The generation of streamlines for one healthy and one tumor subject are shown in Figure~\ref{fig:blockDiagram}(C) and more prominently in Figure~\ref{fig:streamlines}. As seen from the figure, the absence of streamlines is evident within the tumor affected region. This observation is indicative of the disruption of WM tracts within the brain due to the presence of the tumor. Figure~\ref{fig:blockDiagram}(D) displays the resulting connectome, illustrating the number of streamlines connecting 84 DK atlas regions in the brain. Increased structural connectivity within each hemisphere is noticeable and depicted in the diagonal boxes (bright yellow spots), indicating efficient communication between neighboring brain regions. Additionally, brighter spots in the off-diagonal boxes indicate increased structural connectivity between homologous regions of the hemispheres which is essential for various cognitive and motor functions. 
Thus, the connectome serves as a valuable tool for examining how these interactions are altered in the presence of pathology. By mapping the structural connectivity patterns disrupted by the tumor, the connectome helps elucidate the impact of the pathology on the brain's structural and functional organization.

\subsection{Persistent homology based analysis}

\begin{figure}
    \centering
    \includegraphics[width=\textwidth]{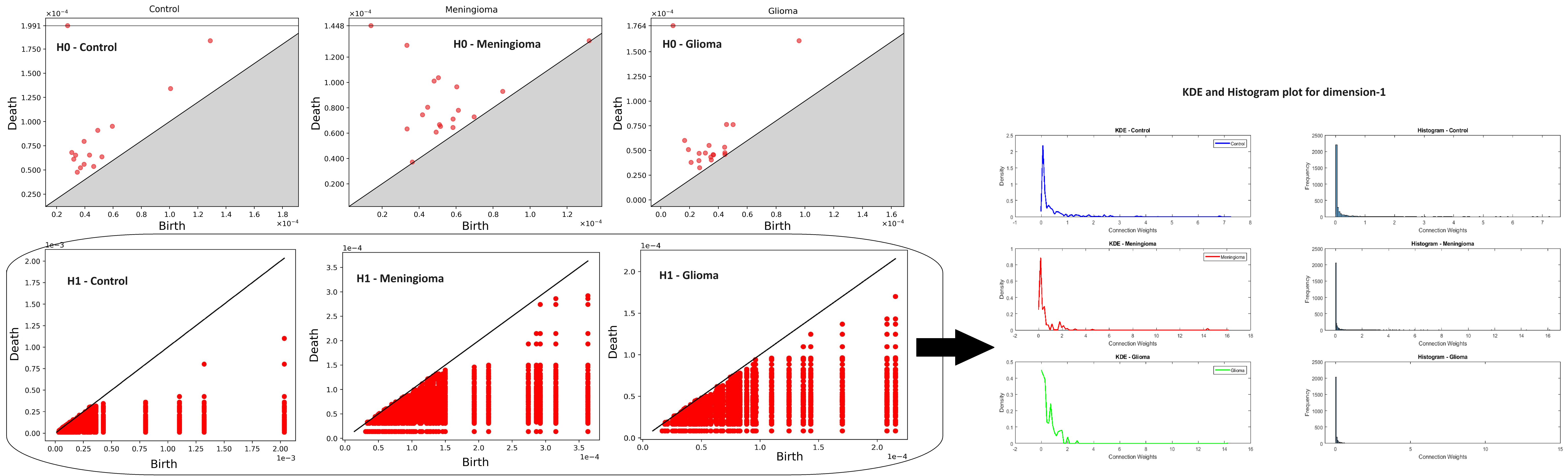}
    \caption{The persistence diagram in dimension-0 (top row in left panel) and dimension-1 (bottom row in left panel) as obtained from whole brain connectome, for one representative subject of control (first column), glioma (second column) and meningioma (third column). The kernel density estimate (KDE) plot and the corresponding histogram for the dimension-1 persistence diagram is shown at the right panel.}
    \label{fig:pd}
\end{figure}

The entire brain connectome is used to compute persistent diagrams, illustrated in Figure~\ref{fig:pd}. The persistence diagrams reveal clear differences in topological features among the groups: Control, Meningioma, and Glioma, for both dimension-0 and -1.
In the control group, a high density of points near the origin in dimension-1 indicates that many loops are born and die quickly. 
In contrast, the tumor groups exhibit points that are more dispersed and away from diagonal line, indicating a wider range of feature persistence. This suggests that the tumor groups have more varied and potentially significant topological features that persist over a broader range of filtration values, likely due to the more complex and disrupted network structure- reflecting the underlying pathological changes in the brain network structure due to presence of tumor.

\begin{figure}
    \centering
    \includegraphics[width=\textwidth]{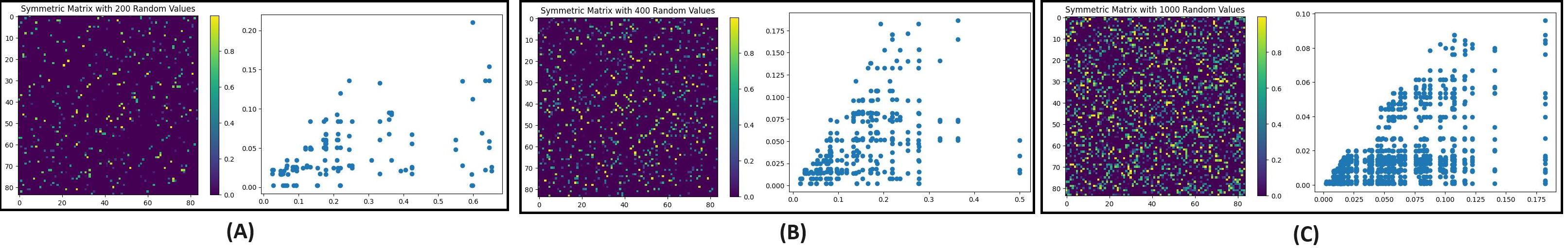}
    \caption{Illustration of the effect of sparsity with synethetic dataset: There is a noticeable increase in the alignment of points along both vertical and horizontal axes in persistence diagram as the level of sparsity is decreased [Fig. (A) to Fig (C)]. 
    The leftmost matrix contains 200 random edge connectivity values (c=200), making it highly sparse. The corresponding persistence diagram is highly scattered, indicating that in a sparse matrix, many features are created and die over a wide range of filtration values. As the sparsity decreases (middle column with c=400), the corresponding persistence diagram shows less scattering compared to the sparse matrix but still maintains a degree of spread. There is a noticeable increase in the alignment of points along both the vertical and horizontal axes as When the sparsity is further decreased, resulting in a denser matrix (right column with c=1000), the corresponding persistence diagram shows many points aligned both horizontally and vertically. This alignment indicates that many features in the topological filtration are born and die at the same or similar filtration values.
 }
    \label{fig:toydata}
\end{figure}

\begin{figure}
    \centering
    \includegraphics[width=\textwidth]{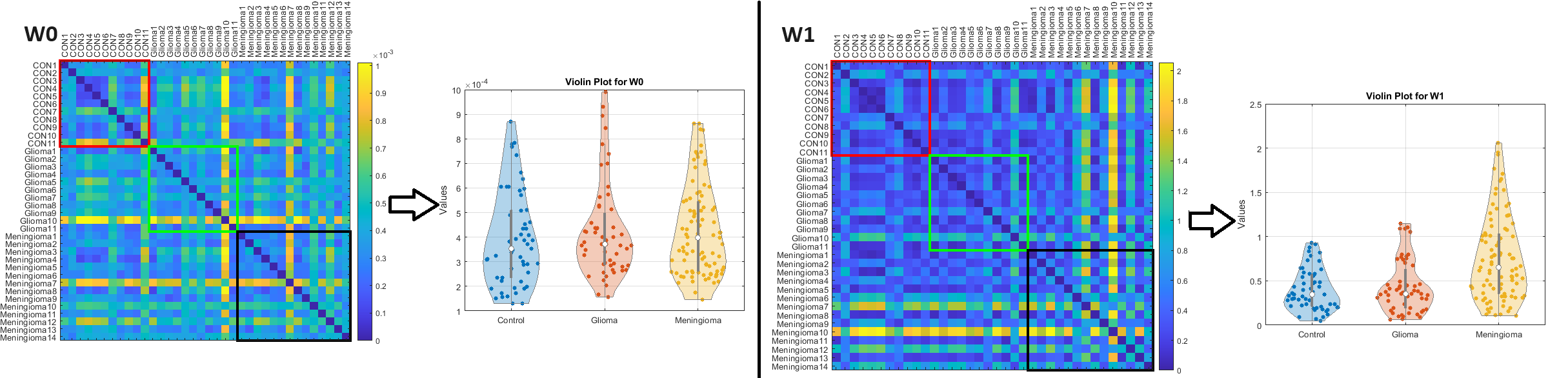}
    \caption{Visualization of Wasserstein distance and the corresponding violin plot for dimension-0 (first and second column) and -1 (third and fourth column), across all subjects to find the statistical significance($p<0.05$) across study groups.}
    \label{fig:wass_violin}
\end{figure}

\subsubsection{Block-type pattern in persistence diagram of brain connectome for $H_{1}$}
As seen from Figure~\ref{fig:pd}, the dimension-1 persistence diagram displays a block-type structure due to several features being born and dying at similar filtration steps. Generally, this is seen to occur under specific conditions:\\
\textit{(i) Dense connectivity or full connected network:} In a dense matrix or fully connected graph, many entries in the matrix may surpass a given threshold simultaneously. As a consequence, multiple features can be created or destroyed at the same filtration step. This can appear as either vertical or horizontal alignments in the persistence diagram, depending on whether features are born or die at similar filtration values. \\
\textit{(ii) Clusters in data:} When non-zero values in the connectivity matrix are concentrated around certain levels, during the filtration process features tend to be created or destroyed at these clustered values. This results in multiple features being born or dying at the same filtration values, leading to aligned points in the persistence diagram.

\textbf{\textit{Proof-of-concept:}} The Figure~\ref{fig:toydata} illustrates condition-(i)- the effect of matrix sparsity on the corresponding persistence diagrams; how the structure and pattern of the connectivity matrix directly influences the topological features and resulting in the observed block-type patterns in persistence diagram.
Figure~\ref{fig:toydata}, depicts how the pattern of persistence diagrams changes with varying degrees of sparsity induced in synthetic symmetric matrices.
As seen from the figure, in a sparse matrix (first column in Fig.A), persistence diagram is scattered reflecting the lack of dominant structures in the distribution of connection weights in the data. In such case, features are born and die over a broad range of filtration values. As the matrix becomes denser (third and fifth columns in Fig.B and Fig.C), it leads to features with similar birth and death values. This results in more structured persistence diagrams, characterized by vertical and horizontal alignments. In context of brain connectivity, the connectome matrix shows clear structure and clustered patterns (Figure~\ref{fig:streamlines}), leading to block-type pattern in persistence diagrams for dimension-1, as shown Figure~\ref{fig:pd}. 
Further explanation to illustrate condition-(ii) is provided using the kernel density estimate (KDE) and the corresponding histogram of brain connectome for one representative subject each from the HC, Meningioma, and Glioma groups in dimension-1, is shown at the right panel in Figure~\ref{fig:pd}. The KDE plots are found to be highly skewed for all subjects. As seen from figure, the first bin of histogram has a very high count. This suggests that many connections have very low weights (zero or nearly zero), which is common in brain connectome matrices as they capture many weak or indirect connections. The high frequency of low weights indicates that multiple features low weights are born at these low filtration values, resulting in vertical alignments in the persistence diagram. As the connection weight or filtration value increases, the frequency of low weights decreases, which is reflected in the gradual reduction of bin counts. Consequently, there are fewer strong connections as compared to weak ones. These histogram bins indicate clusters of connections with similar weights. This clustering implies that many features formed by these stronger connections appeared and disappeared at the similar filtration values, leading to alignments in the persistence diagram.

\subsubsection{Quantification using Wasserstein Distance}
Wasserstein distance is used to quantify the dissimilarities in persistence diagrams between groups. This is shown in Figure~\ref{fig:wass_violin}. Wilcoxon rank-sum test was conducted at 95\% C.I. to determine if the differences in Wasserstein distance are statistically significant between study groups. For dimension-1, the Wasserstein distance shows statistically significant differences between HC vs. Meningioma ($p<0.001$) and Glioma vs. Meningioma ($p<0.001$), illustrated in the violin plot in Figure~\ref{fig:wass_violin}. However, for dimension-0, no significant differences are observed between study groups.

\subsection{Graph theory based analysis}
\begin{figure}
    \centering
    \includegraphics[width=\textwidth]{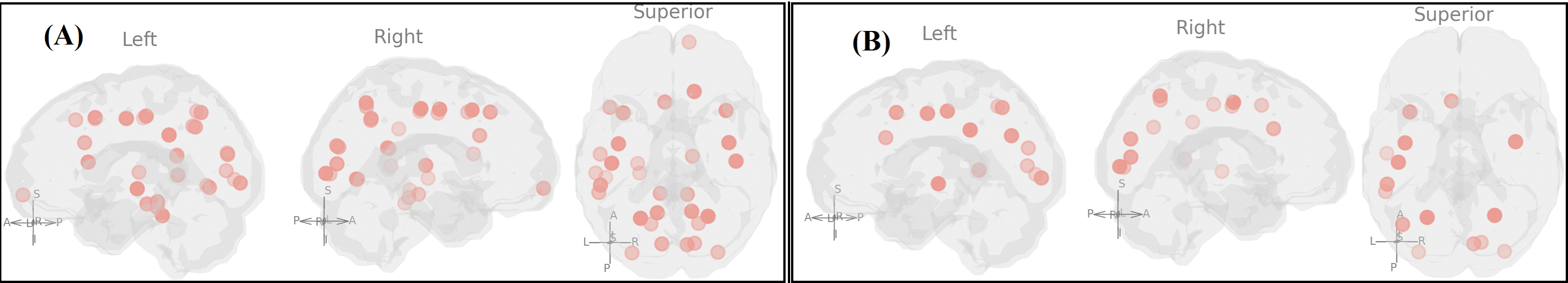}
    \caption{Visualization of DK atlas ROIs where the difference in Local efficiency (Fig.A) and Strength (Fig.B) between HC and Meningioma are found to be statistically significant ($p<0.05$).} 
    \label{fig:visu1}
\end{figure}

\begin{figure}
    \centering
    \includegraphics[width=\textwidth]{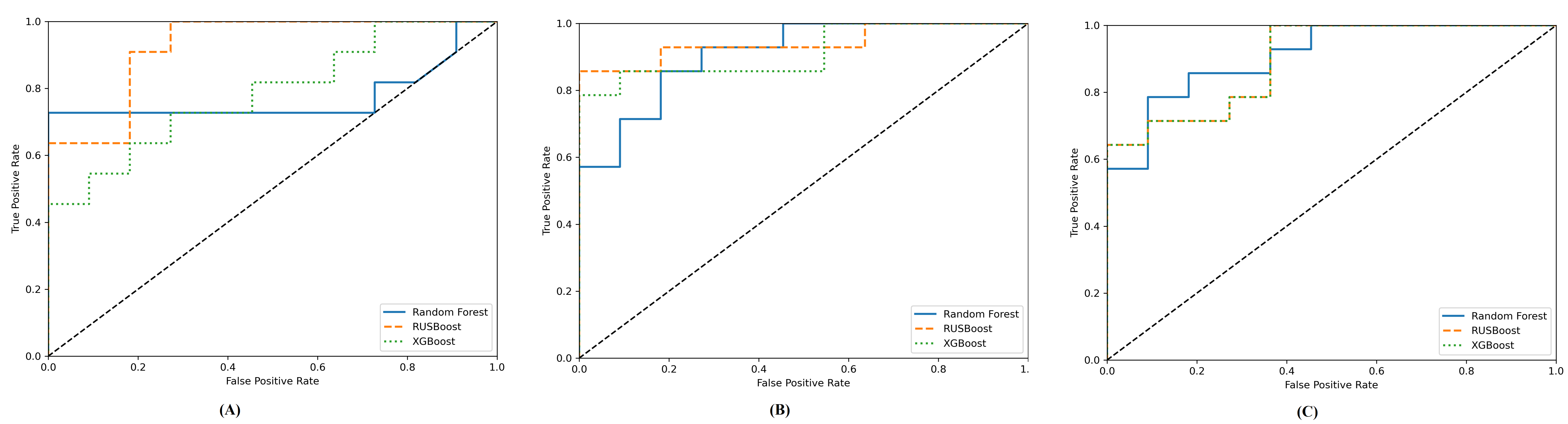}
    \caption{Receiver operating characteristic curves in classifying (A) HC Vs. Glioma (B) HC Vs. Meningioma, and (C) Glioma Vs. Meningioma using whole brain connectome local graph measures. The area under the curve (AUC) for each classifier and for each classification scenarios are illustrated in Table~\ref{tab:localfeatures}.}
    \label{fig:allroi}
\end{figure}

A total of 13 graph-based features from the whole-brain connectome are extracted and utilized as inputs for various machine learning (ML) models to classify different study groups: (i) Glioma Vs. HC, (ii) Meningioma Vs. HC, and (iii) Glioma Vs. Meningioma. For classification purposes, ensemble-based classifiers such as XGBoost, RUSBoost, and Random forest are employed. In the case of using local features for classification, recursive feature elimination is conducted prior to feeding the features into the classifier. However, the classification performance of local features are found to be superior than using global features. The performance of ML classification using both global and local features is presented in Table~\ref{tab:localfeatures}. Due to the smaller sample size, leave-one-out cross-validation is performed.
Among the three different classifiers, the highest accuracies are achieved using RUSBoost, with 88\% for classifying HC Vs. meningioma and 80\% for HC Vs. Glioma. 
In the classification of tumor sub-types, an accuracy of 80\% is achieved in distinguishing meningioma Vs. glioma using RUSBoost and Random forest.
Figure~\ref{fig:allroi} shows the receiver operating characteristic (ROC) curves for each of the three classification scenario. 
Wilcoxon ransum test is conducted at 95\% C.I. to assess the significant differences in local graph feature characteristics for each classification scenarios. This analysis aimed to identify specific brain ROIs where alterations in the graphical properties of WM connectivity are statistically significant. 
Table~\ref{tab:stat} displays the local features indicating significant differences between classes across all subjects. Each element in the table corresponds to the specific ROI in the DK atlas, where the difference in the local feature is statistically significant ($p<0.05$).  The labels of DK ROIs are listed in \textit{Table.2 in supplementary material}.
As seen from Table~\ref{tab:stat}, the statistical analysis reveals numerous ROIs exhibiting significant differences in graph-based measures between meningioma and HC compared to glioma versus HC. Among 9 different local features, local efficiency and strength emerge as the most important features for distinguishing between the two classes. Visualization of significant ROIs for these two features is shown in Figure~\ref{fig:visu1}. However, no significant differences are observed in global graph features between groups.
While numerous studies have achieved comparable accuracy in brain tumor classification \cite{newref2,deb1,deb2,deb3,rajika}, to the best of our knowledge, none have utilized topological features for this task. Our findings are comparable to the insights provided by baseline study \cite{ref5}, highlight the effectiveness of topological data analysis in brain tumor classification. Unlike the baseline study, which focused on functional connectivity prediction and tumor region differentiation without classifying tumor types, our results provide a quantitative measure, demonstrating competitive performance in distinguishing different type of brain tumor by integrating DWI connectomes with persistent homology.

\begin{table}
\begin{center}
\begin{tabular}{|p{1.5cm}|p{0.5cm}|p{0.8cm}|p{0.9cm}|p{2.6cm}|p{0.7cm}|p{0.8cm}|p{0.8cm}|p{0.9cm}|p{1.7cm}|}
\hline
& CC & Deg. & BC & LE & EVC & PC & DC & GC & Str. \\
\hline\hline
Glioma
Vs. HC & ns & 33,82 & ns & 33 & ns & 3,78 & 14,26 & 3,26,
78 & 30 \\
\hline
Menin-
gioma
Vs. HC & ns & 10,33 & 4,21,
80,84 & 1-4,8-10,12,14,15,
17,21,23,24,28-,
30,40,43,53,58,59,
61,69,70,72,73,
76,77,80 & ns & 24,53,
69,73 & 69 & 24,53,
69,73 & 1-3,7,10,21,
23,28-30,53,
59,69,72,77 \\
\hline
Menin-
gioma
Vs. Glioma & ns & 9,
82 & 2,13,
18,54,
72,84 & 2,3,9,13,17,22,
23,31,37,39,65,
72,80,84 & ns & 1,22 & ns & 1,9 & 2,23,65,78 \\
\hline
\end{tabular}
\end{center}
\caption{Results of the statistical analysis, aimed to identify specific brain ROIs where changes in the local graphical properties of DWI brain connectome are statistically significant in relation to glioma and meningioma. Each entry in the table corresponds to particular ROI in the Desikan-Killiany atlas, indicating ROIs where this difference is statistically significant ($p<0.05$). 
In the table,`ns' denotes not statistically significant findings.
}
\label{tab:stat}
\end{table}

\begin{table}
\begin{center}
\begin{tabular}{|c|c|c||c|c|c|c|c|c|}
\hline
Feature &Classifier & Subjects       & Precision & Recall & F1-score & Accuracy & AUC \\
\hline\hline
Global & XGBoost & HC Vs Glioma      & 0.46 & 0.55 & 0.50 & 45\% & 0.45 \\
 &   & HC Vs Meningioma    & 0.68 & 0.93 & 0.79 & 72\% & 0.69\\
  &  & Glioma Vs Meningioma & 0.57 & 0.57 & 0.57 & 52\% & 0.51\\
\cline{2-8}
&RUSBoost & HC Vs Glioma     & 0.55 & 0.55 & 0.55 & 56\% & 0.55\\
 &   & HC Vs Meningioma     & 0.68 & 0.93 & 0.79 & 72\% & 0.69\\
&    & Glioma Vs Meningioma & 0.64 & 0.50 & 0.56 & 64\% & 0.57\\
\cline{2-8}
  &Random & HC Vs Glioma & 0.42 & 0.45 & 0.43 & 50\% & 0.41\\
 & Forest   & HC Vs Meningioma      & 0.63 & 0.71 & 0.66 & 68\% & 0.58\\
  &   & Glioma Vs Meningioma  & 0.57 & 0.57 & 0.57 & 60\% & 0.51\\
\hline
Local &XGBoost & HC Vs Glioma      & 0.78 & 0.64 & 0.70 & \textbf{73\%} & \textbf{0.79} \\
  &   & HC Vs Meningioma     & 0.86 & 0.86 & 0.86 & \textbf{84\%} & \textbf{0.92}\\
  &   & Glioma Vs Meningioma & 0.73 & 0.79 & 0.76 & \textbf{72\%} & \textbf{0.90}\\
\cline{2-8}
&RUSBoost & HC Vs Glioma     & 0.80 & 0.78 & 0.76  & \textbf{77\%} & \textbf{0.93}\\
  &   & HC Vs Meningioma     & 0.92 & 0.86 & 0.89  & \textbf{88\%} & \textbf{0.94}\\
  &   & Glioma Vs Meningioma & 0.85 & 0.79 & 0.81  & \textbf{80\%} & \textbf{0.90}\\
\cline{2-8}
  &Random & HC Vs Glioma & 0.78 & 0.64 & 0.70 & \textbf{77\%} & \textbf{0.81}\\
  & Forest   & HC Vs Meningioma      & 0.81 & 0.93 & 0.87 & \textbf{84\%} & \textbf{0.91}\\
  &   & Glioma Vs Meningioma  & 0.85 & 0.79 & 0.81 & \textbf{80\%} &\textbf{0.91}\\
\hline
\end{tabular}
\end{center}
\caption{Classification performance using graph-based global and local features}
\label{tab:localfeatures}
\end{table}

\section{Discussion}

The present study explores the application of persistent homology and graph-based analysis of DWI brain connectome to differentiate between various types of brain tumors. In the past two decades, numerous imaging studies have underscored the potential of machine learning and deep learning techniques for classifying gliomas using conventional MRI and advanced MRI modalities \cite{glioma_manish9,gliomaclassification_manish1,glioma_manish7,glioma_manish8,deb1,deb2,deb3,rajika}. For example, in \cite{gliomaclassification_manish1}, a custom deep learning model was developed and compared against pretrained models for predicting glioma grades using MRI data, demonstrating superior performance of the custom model and highlighting the effectiveness of deep CNNs and transfer learning in medical image analysis. Other studies have explored deep radiomics features \cite{glioma_manish7,glioma_manish8} and mutational status prediction \cite{deb1,deb2,deb3} in different grades of gliomas, while \cite{rajika} assessed glioma tissue heterogeneity using diffusion tensor and diffusion kurtosis imaging.
Some studies have employed connectome-based predictive modeling to predict individual differences in behavior \cite{connectome_manish2}, trait anxiety in healthy individuals \cite{connectome_manish3}, assessing and managing cognitive decline in aging populations \cite{connectome_manish5} and general intelligence \cite{connectome_manish4}, the primary focus here is on leveraging these advanced techniques to quantify the differences in brain tumors having different origins.

In contrast, the current study adopts a unique and distinct approach that investigates the application of both DWI connectome-based graph measures and topological measures using persistent homology in order to differentiate brain tumors having different origins. 
The significance of this study lies in its ability to provide a comprehensive understanding of the brain's structural connectivity alterations associated with different types of brain tumor. By analyzing connectome data, which represents the connections strength in WM fibre tracks between brain regions, clinicians and researchers can gain insights into the underlying pathology of different types of brain tumor. Connectome-based graph-theoretical approaches offer the advantage of capturing both global and local alterations in brain connectivity, allowing for a more nuanced understanding of tumor-related changes across the entire brain. 
The incorporation of persistent homology adds a layer of depth to the study by capturing and quantifying subtle topological characteristics of the brain connectome. This method identifies topological features in the brain network that persist across different spatial scales, offering a more comprehensive view of how brain connectivity is altered in the presence of meningiomas versus gliomas. This analysis not only enhances the differentiation between tumor types but also contributes to our understanding of the complex network changes induced by these tumors. While further studies are necessary for broader applicability and conclusive inferences, the holistic view offered by connectome analysis in this study has potential to enhance personalized medicine strategies, thereby improving patient outcomes and quality of life. 
The findings from this study could potentially translate into clinical practice by providing neurosurgeons and oncologists with more precise diagnostic tools. The ability to accurately classify tumors having different origin, based on their unique topological signatures may guide surgical planning, treatment selection, and monitoring of treatment response.

\subsubsection{\discintname}
The authors have no competing interests.

\bibliographystyle{splncs04}
\bibliography{ourbib}

\end{document}